\documentclass{svjour3}     
\smartqed  

\usepackage{amsmath}
\usepackage{graphicx}
\usepackage{color}
\usepackage{graphicx}

\newcommand{\be}{\begin{equation}}
\newcommand{\ee}{\end{equation}}
\newcommand{\bea}{\begin{eqnarray}}
\newcommand{\eea}{\end{eqnarray}}

\newcommand{\eref}[1]{Eq.~(\ref{#1})}%
\newcommand{\fref}[1]{Fig.~\ref{#1}} %
\newcommand{\Fref}[1]{Figure~\ref{#1}}%
\newcommand{\sref}[1]{Sec.~\ref{#1}}%
\newcommand{\Sref}[1]{Section~\ref{#1}}%
\newcommand{\tref}[1]{Table~\ref{#1}}%

\begin{document}

\title{Blast waves in two and three dimensions: Euler versus Navier Stokes equations}
\titlerunning{Blast waves}
\author{Amit Kumar   \and R. Rajesh}

\institute{Amit Kumar \at
              	The Institute of Mathematical Sciences, CIT Campus, Taramani, Chennai 600113, India \\
              	Homi Bhabha National Institute, Training School Complex, Anushakti Nagar, Mumbai 400094, India\\
               	\email{kamit@imsc.res.in}
          \and
          	R. Rajesh \at
		The Institute of Mathematical Sciences, CIT Campus, Taramani, Chennai 600113, India \\
              	Homi Bhabha National Institute, Training School Complex, Anushakti Nagar, Mumbai 400094, India\\
               	\email{rrajesh@imsc.res.in}
}

\date{Received: \today / Accepted: }

\maketitle

\begin{abstract}
The exact solution of the Euler equation, which describes the time evolution of a blast wave created by an intense explosion, is a classic problem in gas dynamics. However, it has been found that the analytical results do not match with results from molecular dynamics simulation of hard spheres in two and three dimensions. In this paper, we show that the mismatch between theory and simulations can be resolved by considering the Navier Stokes equation.  From the direct numerical simulation of the Navier Stokes equation in two and three dimensions, we show that the inclusion of heat conduction and viscosity terms is essential to capture the results from molecular dynamics simulations.
\end{abstract}

\section{\label{sec1-Introduction}Introduction}

The time evolution of a blast wave resulting from an intense explosion  is a classic problem in gas dynamics~\cite{landaubook,barenblatt1996scaling,whitham2011linear}. After an initial transient period when energy is transported in the form of radiation,  the system enters a hydrodynamic regime where radiation becomes less important, and the  transport of energy is dominated by transfer of matter. The disturbance grows radially outwards and a shock front forms the boundary between  the perturbed gas and the ambient gas. Across the shock front, thermodynamic quantities like density, velocity, temperature, and pressure are known to be discontinuous~\cite{landaubook,whitham2011linear}. 

From dimensional analysis, the radius of the shock front $R(t)$ is known to increase with time $t$ as a power law $R(t)\sim (E_0t^2/\rho_0)^{1/(d+2)}$ in $d$ dimensions, where $E_0$ is the energy of the blast and $\rho_0$ is the ambient density of the system~\cite{taylor1950formation,taylor1950formation2,jvneumann1963cw,sedov_book,sedov1946}. This scaling law has been verified in many different contexts.  The earliest and an accurate measurement was from the data of the Trinity nuclear explosion~\cite{taylor1950formation,taylor1950formation2}. Later on, this scaling law has been verified experimentally in laser-driven blast waves in gas jets~\cite{edwards2001investigation}, plasma~\cite{edens2004study}, or in atomic clusters of different gases~\cite{moore2005tailored}.

Theoretically, in addition to $R(t)$, it has also been possible to obtain  the spatio-temporal behavior of different thermodynamic quantities in the hydrodynamic regime. The variation of density, pressure, temperature, and velocity with spatial distance, $r$, and time, $t$, are described by continuity equations for mass, momentum, and energy. These hydrodynamic equations correspond to the Navier Stokes equation. In the scaling limit $r \to \infty $, $t \to \infty$ keeping $r t^{-2/(d+2)}$ fixed, and after appropriately non-dimensionalizing the thermodynamic quantities, it was shown that the heat conduction and viscosity terms vanish and hence can be dropped. The Euler equations, resulting from dropping heat conduction and viscosity terms, in three dimensions were solved exactly by Taylor, von-Neumann and Sedov~\cite{taylor1950formation,taylor1950formation2,jvneumann1963cw,sedov_book,sedov1946} to obtain the scaling functions analytically, and we will refer to these self-similar solutions as the TvNS solution. The significance  of the TvNS solution is in its wide applicability. For instance, it has been used for modeling evolution of supernova remnants in the adiabatic stage~\cite{gull1973numerical,cioffi1988dynamics,ostriker1988astrophysical,cowie1977early,bertschinger1985cosmological,bertschinger1983cosmological}. TvNS theory has been used to study systems where energy is continuously input in a localized region of space~\cite{dokuchaev2002self,falle1975numerical}.
TvNS theory has  also been generalized for granular systems, where conserved quantities are fewer in number because energy is no longer conserved~\cite{barbier2015blast,barbier2016microscopic}. There are many examples in granular systems where a blast wave is generated when perturbed by either a single impact or by continuous energy injection. For instance, crater formation in granular bed due to single impact of an object or continuous jet~\cite{walsh2003morphology,metzger2009craters,grasselli2001crater}, shock propagation in granular medium due to sudden impact~\cite{boudet2009blast,jabeen2010universal,pathak2012shock}, viscous fingering due to continuous energy injection~\cite{cheng2008towards,sandnes2007labyrinth,pinto2007granular,johnsen2006pattern,huang2012granular} or shock propagation in continuously driven granular media~\cite{joy2017shock}.  

Unlike the scaling law for $R(t)$, direct  experimental determination of the scaling functions for density, temperature, velocity and pressure is limited to measurement of the density profile for one single time (see Fig.~1 of \cite{edwards2001investigation} and inset of Fig.~5 of \cite{moore2005tailored}). However, these data are insufficient to make a meaningful comparison with the theoretical predictions for the scaling functions from the TvNS theory. In the absence of experimental data, molecular dynamics (MD) simulations of hard spheres can serve as a model platform for verifying the TvNS solution.

Recently, the TvNS theory has been tested and compared with MD simulations of hard spheres in two and three dimensions.  Initial MD studies verified the  power-law growth of the radius of the shock front in two~\cite{jabeen2010universal,antal2008exciting} and three dimensions~\cite{jabeen2010universal}, consistent with the TvNS theory. Later work focused on verifying the theoretical predictions of the scaling functions for density, radial velocity, temperature and pressure. Among these, one study of the scaling functions  in two dimensions concluded that the TvNS theory describes well the simulation data  for low to medium number densities except for a small difference near the shock front and slight discrepancy near the shock center~\cite{barbier2016microscopic}. However, by performing more extensive and large scale simulations, we found that  the numerically obtained radial distribution of density, radial velocity as well as the temperature  do not match with the TvNS theory (modified to account for steric effects) neither in two dimensions~\cite{joy2021shock} nor in three dimensions~\cite{joy2021shock3d}. The differences were manifested in different power law behavior at the shock center for density, radial velocity and temperature even though the pressure distributions from theory and simulations matched. 

An explanation that we postulated for the origin of the mismatch between MD simulations and TvNS solutions was the non-commutability of the order of taking limits: the solution of the Euler equations after taking the scaling limit (as in TvNS solution) is not the same as applying the scaling limit to the solution of the Navier Stokes  equations that include heat conduction and viscosity.
In particular, heat conduction terms impose the boundary condition of $\nabla T=0$ at $r=0$, $T$ being the temperature, which the TvNS solution can be seen to not obey~\cite{joy2021shock3d}.  Recently, it was shown~\cite{ganapa2021blast,chakraborti2021blast} that  the results from direct numerical simulations (DNS) of the Navier Stokes equations, which include heat conduction and viscosity terms,  compared very well with the results from MD simulations of shock propagation in a one dimensional gas consisting of two kinds of point particles. These studies also argue that the heat conduction term becomes important at distances very close to the shock center.

In this paper, we ask if Navier Stokes equation will provide the correct description of the data from MD simulations in two and three dimensions. To do so, we do DNS of the continuity equations for mass, momentum and energy using the same numerical methods followed in Ref.~\cite{ganapa2021blast}. In continuation of our earlier work~\cite{joy2021shock,joy2021shock3d}, steric effects, which are absent in one-dimensional systems, were included by modifying the equation of state that relates pressure to temperature and density from the ideal gas equation of state to the virial equation of state.
The boundary conditions for the equations in radial coordinates are obtained from  DNS of the equations in cartesian coordinates. The constants related to heat conduction and viscosity are determined from kinetic theory. We then show that the results from DNS match the data from MD simulations without any fitting parameter in two dimensions. In three dimensions, there is a small mismatch which is resolved by treating the constants appearing in heat conduction and viscosity terms as free parameters. We conclude that to obtain the correct description of a blast wave in the hydrodynamic regime, the solution has to be first found for the Navier Stokes equation, which includes heat conduction and viscosity terms, followed by taking the appropriate scaling limit. 

The remainder of the  paper is organized as follows. In \sref{sec1-Hydrodynamics and Numerical details}, we review the continuity equations  to describe hydrodynamics of shock in the spherical polar coordinate in $d$-dimensions. The numerical algorithm and the values of the parameters used for DNS are also given. In \sref{sec3-Results}, we show that the results from DNS match with the results from MD of hard spheres in both two and three dimensions.  \Sref{sec4-Summary and discussion } contains a summary and discussion.

\section{\label{sec1-Hydrodynamics and Numerical details}Hydrodynamics and Numerical details}

\subsection{Hydrodynamics}

We briefly review the hydrodynamics of explosion induced shock propagation in $d$-dimensions. Consider a gas of constant density $\rho_0$ and temperature zero. At time $t=0$, energy $E_0$ is input isotropically  in a localized region around $r=0$. The state of the gas is described by its density field $\rho(\vec{r},t)$, velocity field $\vec{u}(\vec{r},t)$, temperature field $T(\vec{r},t)$, and pressure field $p(\vec{r},t)$.  Since mass, momentum and energy are locally conserved, their continuity equations describe the time evolution of the local fields.

In spherical polar coordinates, the continuity equations for density, momentum, and energy for a mono-atomic gas along the radial direction are respectively~\cite{landaubook,whitham2011linear,huang1963statistical},
\begin{eqnarray}
&&\partial_t\rho +\frac{1}{r^{d-1}} \partial_r(r^{d-1}\rho u) = 0,\label{eq1}\\
&&\partial_t(\rho u) + \frac{1}{r^{d-1}}\partial_r(r^{d-1}\rho u^2 )+ \partial_r p  = \frac{1}{r^{d-1}}\partial_r(2\mu r^{d-1}\partial_r u) - \frac{2\mu (d-1)u}{r^2}, \label{eq2}\\
&&\partial_t\Big[\frac{1}{2}\rho u^2 + \frac{1}{\gamma-1}\rho T\Big] +  \frac{1}{r^{d-1}}\partial_r\Big[r ^{d-1}\Big(\frac{1}{2}\rho u^2 + \frac{1}{\gamma-1}\rho T + p \Big)u\Big] \nonumber\\
&&\qquad=  \frac{1}{r^{d-1}}\partial_r(2\mu r^{d-1} u\partial_r u) +  \frac{1}{r^{d-1}}\partial_r( r ^{d-1} \lambda \partial_r T) ,\label{eq3}
\end{eqnarray}
where $\gamma = 1+\frac{2}{d}$, $\mu$ is the coefficient of viscosity and $\lambda$ is the coefficient of heat conduction.

The coefficients $\mu$ and $\lambda$ increase with temperature as~\cite{huang1963statistical,reif2009fundamentals} 
\begin{align}
\mu &= C_1 \sqrt{T},\label{eq4}\\
\lambda &=C_2 \sqrt{T}, \label{eq5}
\end{align}
where $C_1$ and $C_2$ are constants which depend on the size of the particles. From  kinetic theory of gases~\cite{reif2009fundamentals}, $C_1$ and $C_2$ can be obtained in $d$-dimensions for particles with diameter $D$ as
\begin{align}
C^*_1 &= \frac{S_d}{2dC^{(d)}_D}\sqrt{\frac{mk_B}{\pi^d}}~  \Gamma\Big(\frac{d+1}{2}\Big),\label{eq6}\\
C^*_2 &= \frac{S_d}{4C^{(d)}_D}\sqrt{\frac{k_B^3}{m\pi^d}}~  \Gamma\Big(\frac{d+1}{2}\Big), \label{eq7}
\end{align}
where $S_d$ is the surface area of a $d$-dimensional sphere of radius one, $C_D^{(d)}$ is the cross sectional area of $d$-dimensional sphere of radius $D$, $m$ is the mass of a particle and $\Gamma$ is the gamma function.  The star in the superscript denotes the kinetic theory result for $C_1$ and $C_2$.

Equations~(\ref{eq1})-(\ref{eq3}) have four independent fields. If local thermal equilibrium is assumed, as is usually done, then the local pressure $p$ is expressed  in terms of local density $\rho$ and local temperature $T$ through an equation of state (EOS). While the ideal gas EOS was used in the original TvNS solution, for a hard sphere gas, steric effects become important and a more realistic EOS would be required in order to compare with results from hard sphere simulations~\cite{joy2021shock,barbier2016microscopic,joy2021shock3d}. We choose the EOS to be the virial EOS, which takes the form
\begin{align}
p &= k_B \rho T \left[1+\sum_{i=2}^\infty B_i\rho^{i-1}\right],\label{eq10}
\end{align}
where $B_i$ is $i^{th}$ virial coefficient. In \tref{table1}, we tabulate the values for the first $10$ virial coefficients. 
\begin{table}
\caption{\label{table1} 
The numerical values of the virial coefficients $B_i$ for the hard sphere gas in two and three dimensions. The data are from Ref.~\cite{mccoy2010advanced},  and for spheres of diameter one. } 
\begin{tabular}{lll}
 \hline\noalign{\smallskip}
$i$  & $B_i(d=2)$ & $B_i(d=3)$  \\[0.5ex]
 \noalign{\smallskip}\hline\noalign{\smallskip}
 $2$ & $\frac{\pi}{2}$ & $\frac{2\pi}{3}$\\
 $3$ & $ (\frac{4}{3}-\frac{\sqrt{3}}{\pi}) B_2^2$ & $\frac{5}{8}B_2^2$\\
 $4$ & $\Big[2-\frac{9 \sqrt{3}}{2 \pi}+\frac{10}{\pi ^2}\Big] B_2^3$ & $\Big[\frac{2707}{4480} + \frac{219\sqrt{2}}{2240\pi} - \frac{4131}{4480}\frac{\arccos[1/3]}{\pi} \Big]B_2^3$\\ 
 $5$ & $0.33355604 B_2^4$ & $0.110252B_2^4$\\
 $6$ & $0.1988425 B_2^5$ & $0.03888198B_2^5$\\
 $7$ & $0.11486728 B_2^6$ & $0.01302354B_2^6$\\
 $8$ & $0.0649930 B_2^7$ & $0.0041832B_2^7$\\
 $9$ & $0.0362193 B_2^8$ & $0.0013094B_2^8$\\
 $10$ & $0.0199537 B_2^9$ & $0.0004035B_2^9$\\[1ex]
 \noalign{\smallskip}\hline
\end{tabular}
\end{table}

\subsection{\label{Numerical method}Numerical method}

We solve  Eqs.~(\ref{eq1})-(\ref{eq3}) numerically using the MacCormack numerical integration method~\cite{maccormack1982numerical}.   
In this scheme, a partial differential equation is solved by  using predictor-corrector steps.  It has accuracy upto second order both in spatial discretisation $\Delta r$ and temporal discretisation $\Delta t$.
\begin{table}
\caption{\label{table2} 
The numerical values of different parameters used in solving Eqs.~(\ref{eq1})-(\ref{eq3}) in two and three dimensions.} 
\begin{tabular}{lll}
 \hline\noalign{\smallskip}
 Parameters & $d=2$  &   $d=3$  \\[0.5ex]
 \noalign{\smallskip}\hline\noalign{\smallskip}
 $\rho_0$ &  $0.382$  &  $0.4$\\
 $L$ & $300$  &  $136$\\
 $\tau$ & $8000$  &  $10000$\\ 
 $\Delta r$ & $0.05$  &  $0.05$\\
 $\Delta t$ & $10^{-5}$  & $10^{-5}$\\
 $\sigma$ & $0.5$    & $0.65$\\
 $E_0$ & $4.0$  & $6.0$\\
 $\gamma$ & $2.0$  & $5/3$\\
 $k_B$ & $1.0$ & $1.0$\\
 $m$ & $1.0$ & $1.0$\\
 $D$ & $1.0$ & $1.0$\\
 $C^*_1$ & $\sqrt{\pi}/8$  & $2/3\sqrt{\pi^3}$\\
 $C^*_2$ & $\sqrt{\pi}/8$  & $1/\sqrt{\pi^3}$\\[1ex]
 \noalign{\smallskip}\hline
\end{tabular}
\end{table} 

We now describe the parameters that we have used for numerically solving  Eqs.~(\ref{eq1})-(\ref{eq3}). We choose system size $L$ and total integration time $\tau$ such that the shock does not reach the boundary at time $\tau$. The initial condition at time $t=0$ is: density $\rho=\rho_0$ everywhere, velocity $u=0$ everywhere and an initial gaussian temperature profile 
\be
T(r,0) = \frac{T_0}{\sqrt{2\pi\sigma^2}}\exp\left(-\frac{r^2}{2\sigma^2} \right),\label{eq11}
\ee
where, for a mono-atomic gas, $T_0=E_0(\gamma-1)/\rho_0$, where $E_0$ is the initial energy given to the system.
In addition to the initial conditions, we also need the boundary conditions at $r=0$ and $r=L$. At $r=0$, we impose the boundary conditions $\partial_r \rho =0$, $u=0$, and $\partial_r T =0$. These boundary conditions are justified by  numerically solving the continuity equations in cartesian coordinates, wherein the boundary conditions at the origin need not be specified. The details are given in appendix~\ref{appendix}, and the results are consistent with the above boundary conditions.  At $r=L$, we use  $\rho=\rho_0$, $u=0$, and $T=0$.

The numerical values of the different parameters are given in \tref{table2}.

\section{\label{sec3-Results}Results}

We define non-dimensionalised density $\widetilde{\rho}$, radial velocity $\widetilde{u}$, temperature $\widetilde{T}$, and pressure $\widetilde{P}$ fields. From dimensional analysis~\cite{barenblatt1996scaling},
\begin{align}
\rho(\vec{r},t)&= \rho_0 \widetilde{\rho}(\xi),\label{eq12}\\
u(\vec{r},t)&=\frac{r}{t} \widetilde{u}(\xi),\label{eq13}\\
T(\vec{r},t)&=\frac{r^2}{t^2}\widetilde{T}(\xi),\label{eq14}\\
p(\vec{r},t)&=\frac{\rho_0 r^2}{t^2} \widetilde{P}(\xi),\label{eq15}
\end{align}
where 
\be
\xi = r \left(\frac{E_0 t^2}{\rho_0}\right)^{-1/(d+2)}, \label{eq16}
\ee
is the scaled distance. We describe the results for two dimensions in \sref{sec:2d} and three dimensions in \sref{sec:3d}.

\subsection{\label{sec:2d} Results in two dimensions}

\begin{figure}
\centering
\includegraphics[width=0.8\columnwidth]{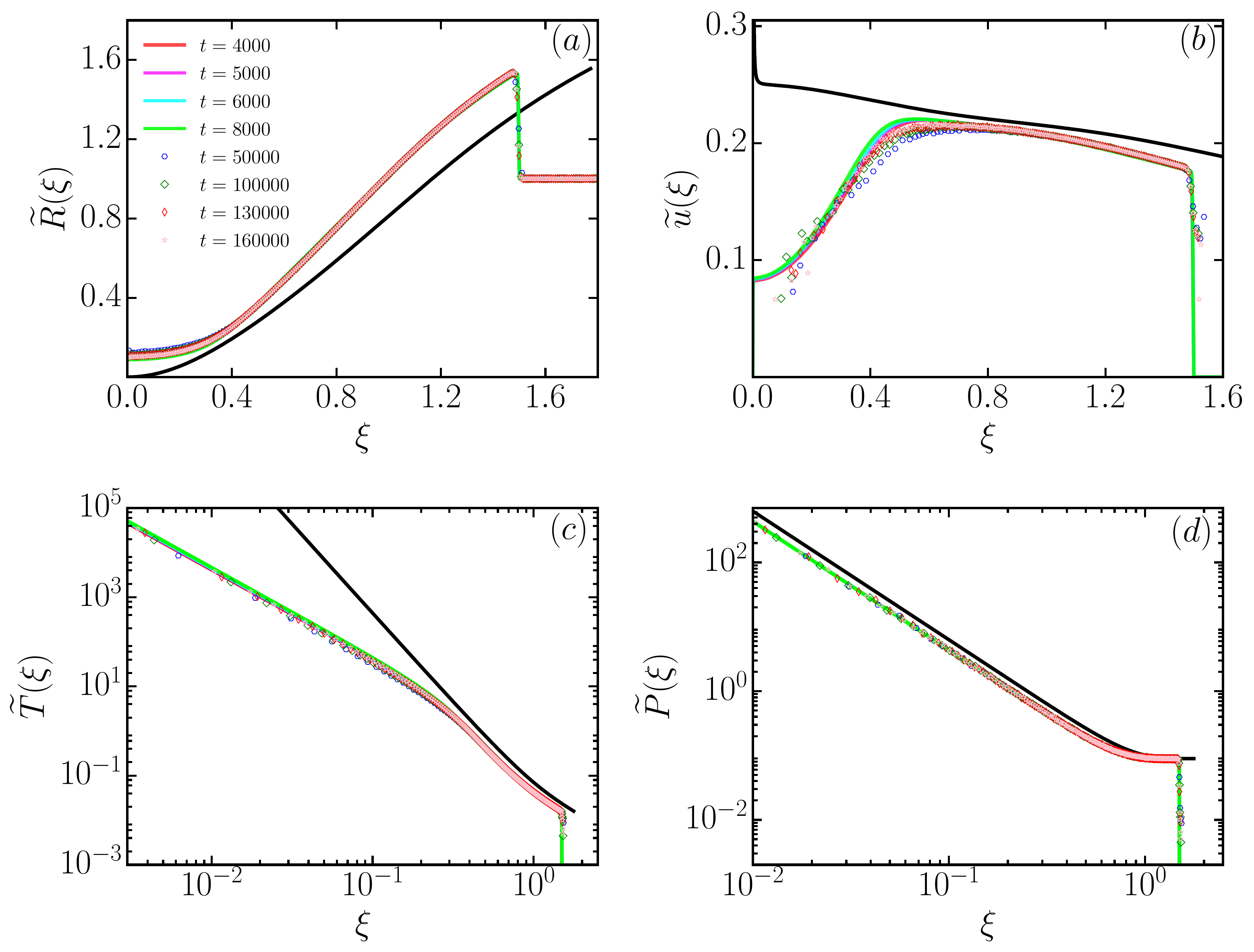}
\caption{The data from MD simulations (taken from Ref.~\cite{joy2021shock}) are compared with the data from DNS of the continuity Eqs.~(\ref{eq1})-(\ref{eq3}) in two dimensions. The variation of the non-dimensionalised (a) density (b) radial velocity (c) temperature and (d) pressure with $\xi$ is shown for four different times (different times for MD and DNS).  The symbols represent the MD data and colored lines represent the results from DNS.  The solid black lines represent the TvNS solution with virial EOS. The data are for ambient gas density $\rho_0=0.382$.}
\label{fig01}
\end{figure}

We first describe our main result. \Fref{fig01} shows the variation of the different non-dimensionalised observables with $\xi$ for four different times. The data for different times collapse onto one curve confirming the scaling in Eqs.~(\ref{eq12})-(\ref{eq15}). The results from DNS are an excellent match with the results from MD simulations (the MD data are from Ref.~\cite{joy2021shock}) for all the four thermodynamic quantities.  In comparison, the TvNS solution, shown by solid black lines, does not describe the data well. We note that there are no fitting parameters used in the simulations, with the values $C_1^*$ and $C_2^*$ being calculated from kinetic theory. We conclude that heat conduction and viscosity terms are essential to reproduce the results from MD, though these terms are irrelevant in the scaling limit.

\begin{figure}
\centering
\includegraphics[width=0.8\columnwidth]{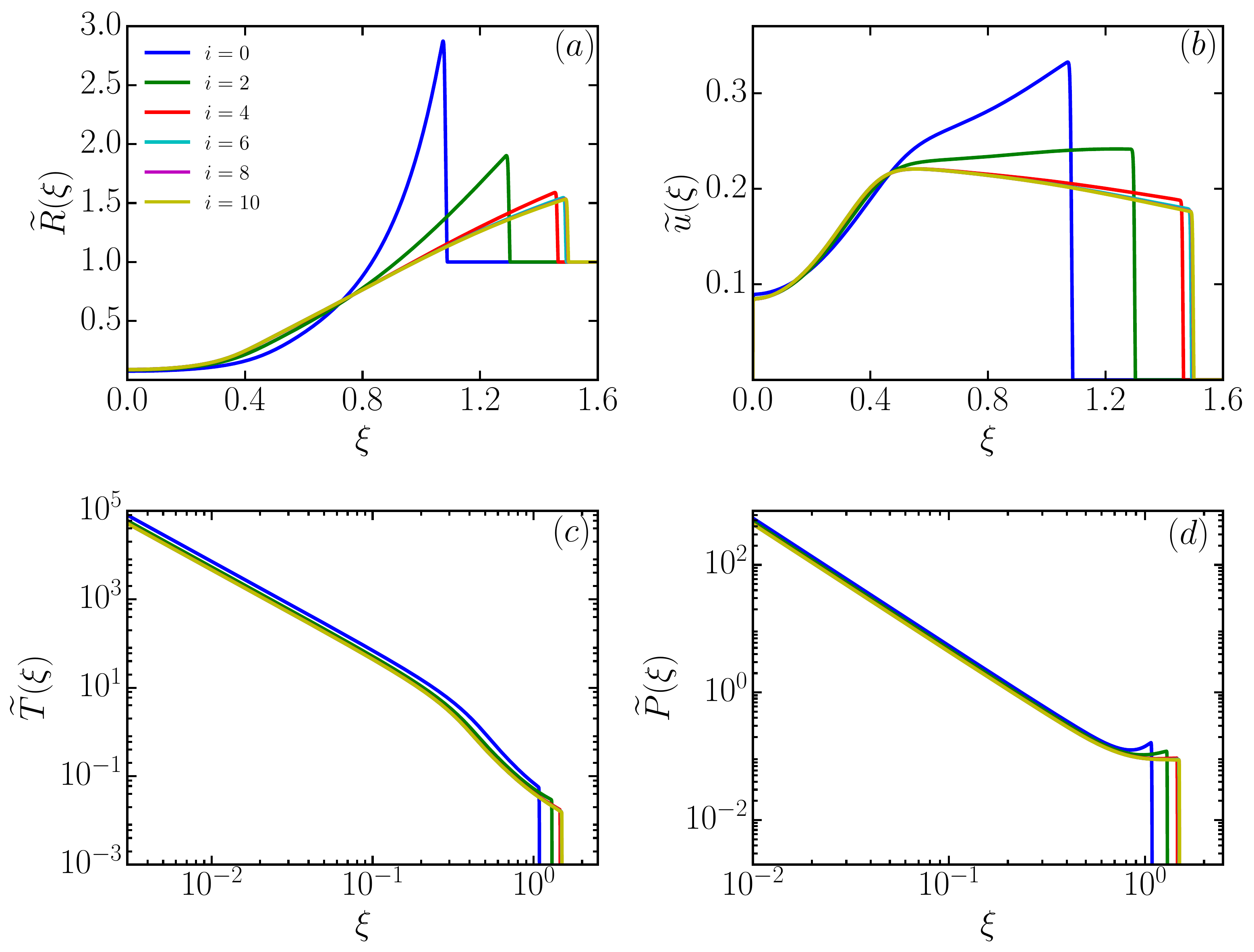}
\caption{The variation of the non-dimensionalised (a) density $\rho(r)$, (b) velocity $u(r)$,  (c) temperature $T(r)$, (d) pressure $p(r)$ obtained from the DNS of the continuity Eqs.~(\ref{eq1})-(\ref{eq3}) in two dimensions, when  the virial expansion in \eref{eq10} is truncated at $i=2,4,6,8,10$.  The data for different $i$ are shown for a single time $t=8000$ and ambient density $\rho_0=0.382$.  $i=0$ corresponds to ideal EOS $p=\rho k_B T$. }
\label{fig02}
\end{figure}

We have used the virial EOS [see \eref{eq10}] for relating local pressure to density and temperature. However, only ten terms of the expansion are known. To check whether our numerical results are affected by this truncation, we solve the continuity equations by using the virial EOS only upto the $i^{th}$ term where $i=0, 2,4, \ldots, 10$. $i=0$ corresponds to the ideal gas. \Fref{fig02} shows the dependence of our results on the number of terms kept in the virial expansion. The results for $i=8$ and $i=10$ are nearly identical, showing that truncating the virial EOS at the $10^{th}$ term causes negligible error in the results.

\begin{figure}
\centering
\includegraphics[width=0.8\columnwidth]{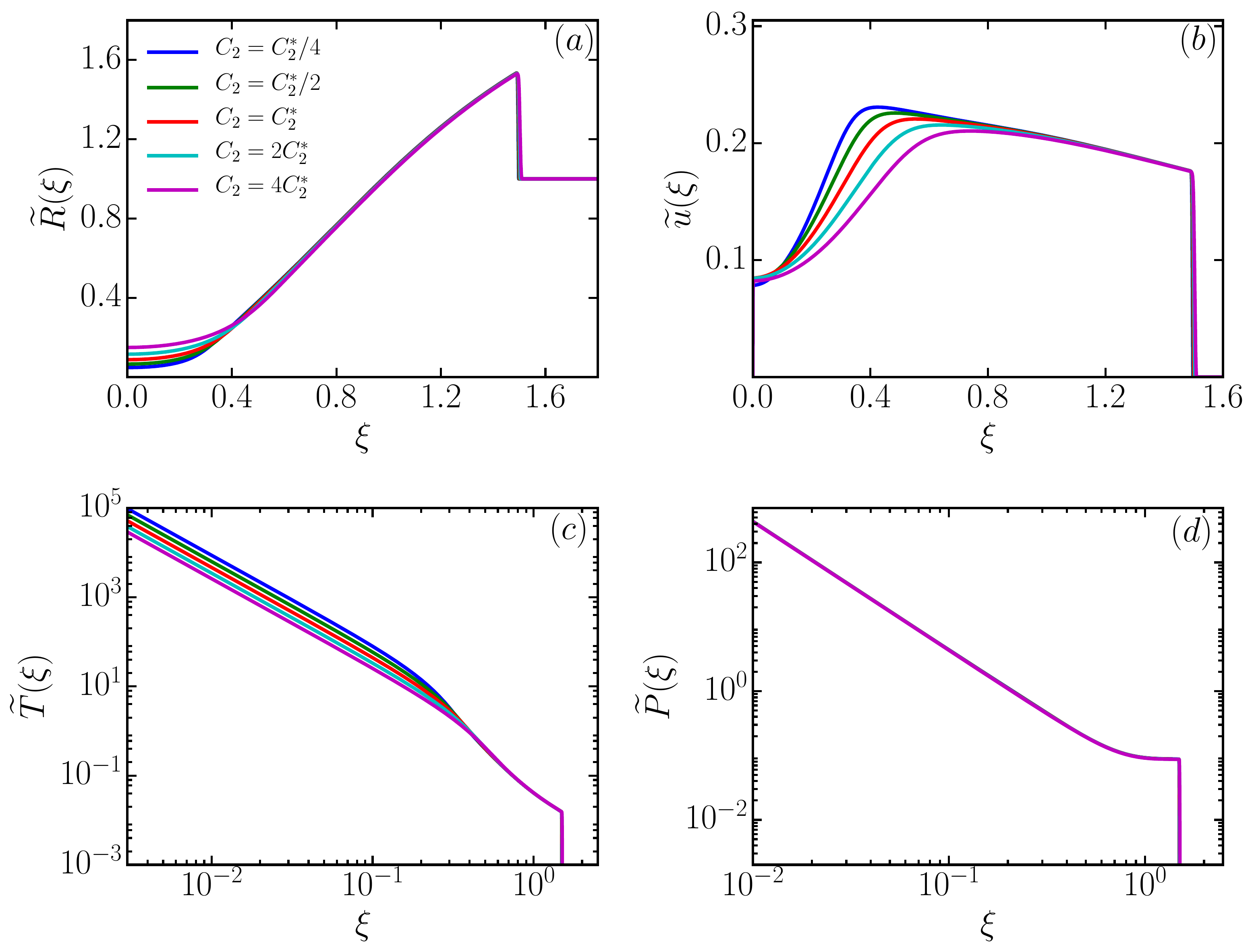}
\caption{Non-dimensionalised (a) density, (b) radial velocity,  (c) temperature, and  (d) pressure for five different values of the parameter $C_2$ [see \eref{eq5}] keeping all other parameters fixed. The data are for  $t=8000$ and ambient density $\rho_0=0.382$. } 
\label{fig03}
\end{figure}

We now probe the sensitivity of the results to the parameters $C_1$ and $C_2$ where $\mu = C_1 \sqrt{T}$ and $\lambda =C_2 \sqrt{T}$  [see Eqs.~(\ref{eq4})-(\ref{eq5})]. 
In~\fref{fig03}, we show the results for the different thermodynamic quantities for different values of $C_2$ keeping all other parameters fixed. We choose $C_2=C^*_2/4,C^*_2/2,C^*_2,2C^*_2,4C^*_2$. Near the shock front, these parameters have negligible effect. On the other hand, near the shock center, the results depend on $C_2$. However, features such as the power-law exponents, density having a finite value at $r=0$ remain unchanged with changing $C_2$. We also note that pressure does not have any appreciable dependence on $C_2$ for all $\xi$.

\begin{figure}
\centering
\includegraphics[width=0.8\columnwidth]{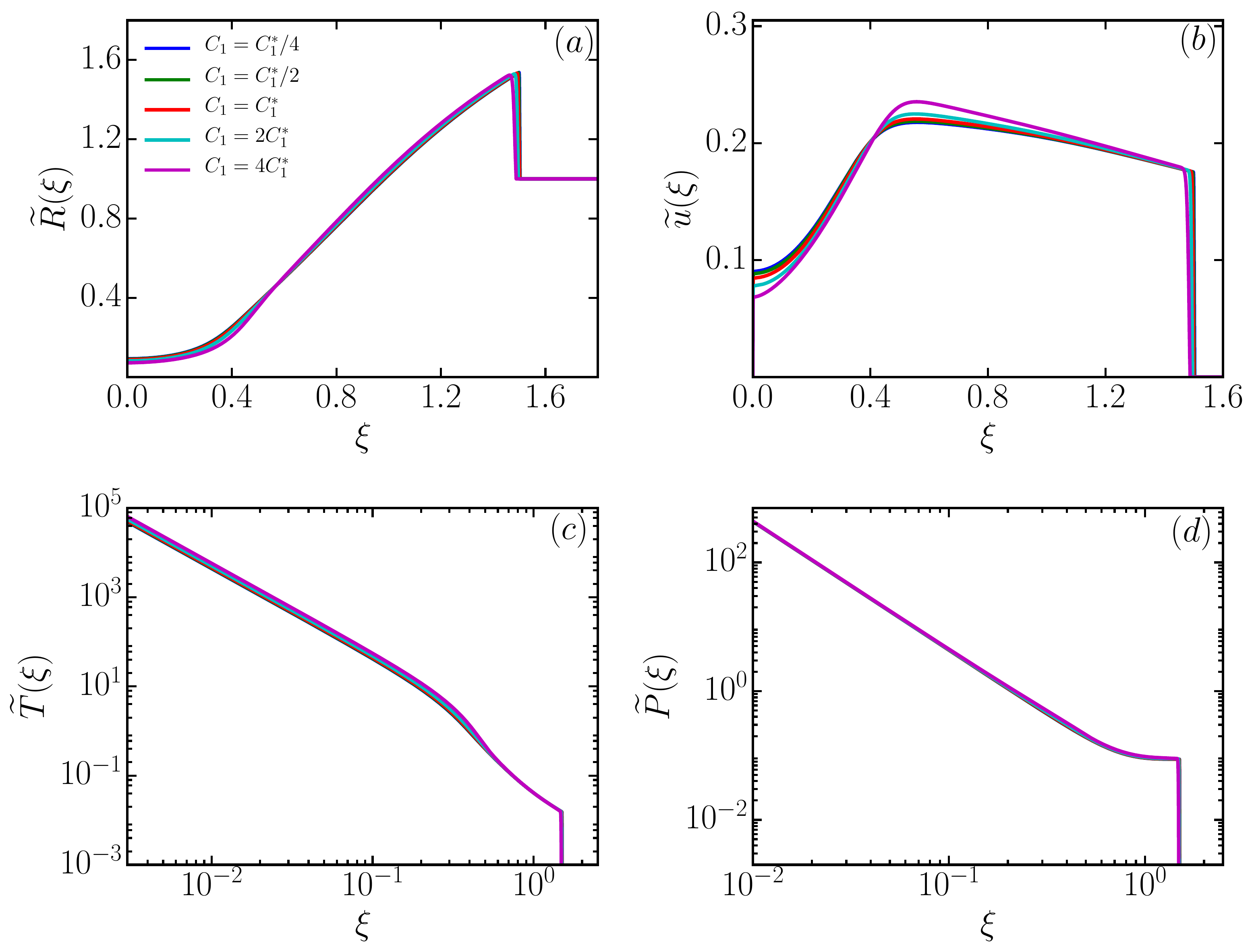}
\caption{ Non-dimensionalised (a) density, (b) radial velocity,  (c) temperature, and  (d) pressure for five different values of the parameter $C_1$ [see \eref{eq4}] keeping all other parameters fixed. The data are for  $t=8000$ and ambient density $\rho_0=0.382$. } 
\label{fig04}
\end{figure}

\Fref{fig04} shows the dependence on the parameter $C_1$, where the results are shown for $C_1=C^*_1/4,C^*_1/2,C^*_1,2C^*_1,4C^*_1$ keeping all other parameters fixed. As for $C_2$, we find that pressure does not depend on $C_1$. For the other quantities, changing $C_1$ does not affect the data near the shock front, while it affects the data near the shock center. However, features such as the power-law exponents, density having a finite value at $r=0$ remain unchanged with changing $C_1$.

\subsection{\label{sec:3d} Results in three dimensions}

\begin{figure}
\centering
\includegraphics[width=\columnwidth]{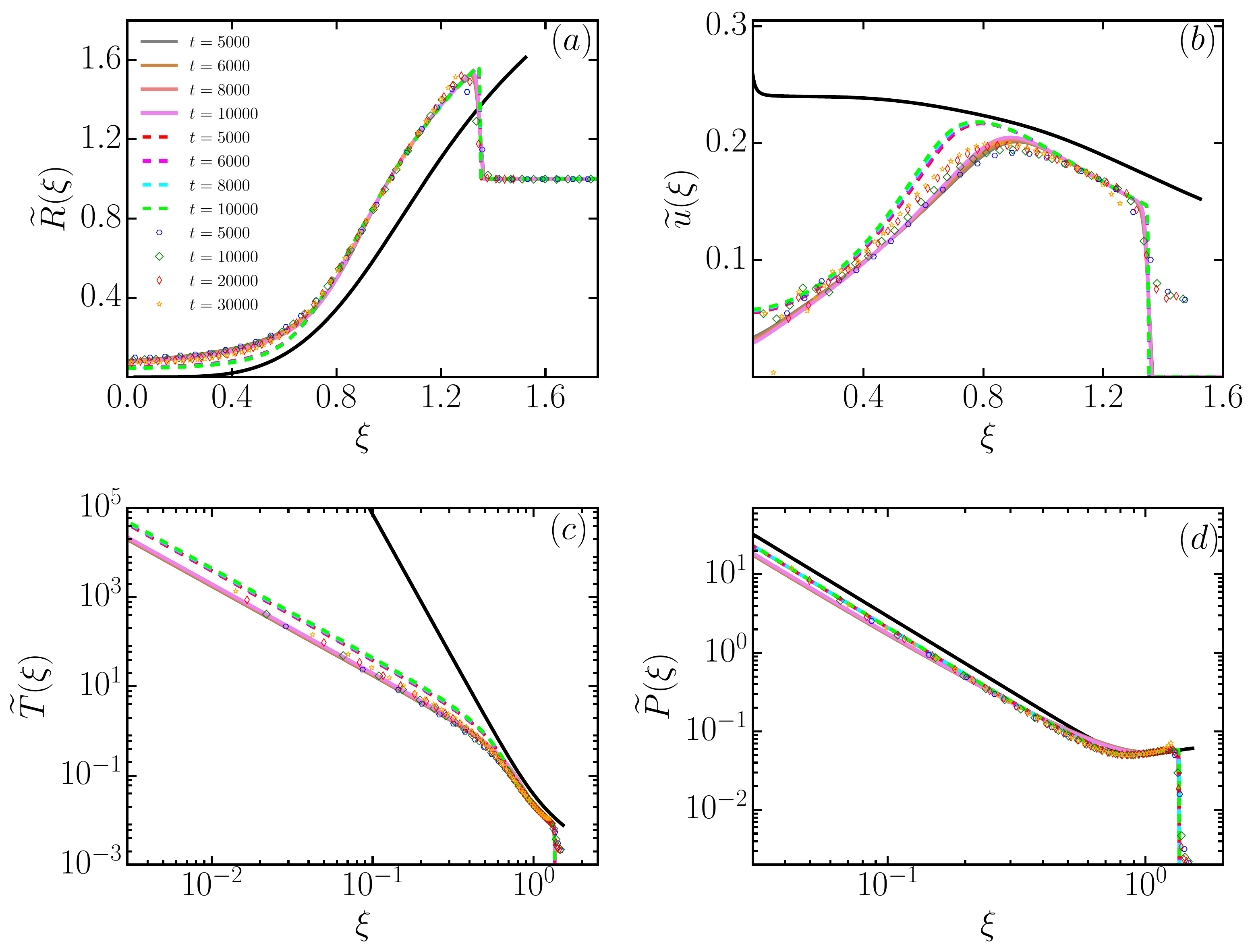}
\caption{The data from MD simulations (taken from Ref.~\cite{joy2021shock3d}) are compared with the data from DNS of the continuity Eqs.~(\ref{eq1})-(\ref{eq3}) in three dimensions. The variation of the non-dimensionalised (a) density (b) radial velocity (c) temperature and (d) pressure with $\xi$ is shown for four different times. The symbols represent the MD data and solid colored lines represent the DNS results of the continuity Eqs.~(\ref{eq1})-(\ref{eq3}) with $C_1=2/3$, $C_2=1$, while the dashed lines represent the DNS results for $C_1=C_1^*$ and $C_2=C_2^*$. The solid black lines represent the TvNS solution with virial EOS. The data are for ambient gas density $\rho_0=0.4$.}
\label{fig05}
\end{figure}

We now show that the Navier Stokes equation in three dimensions is able to reproduce the simulation results from MD.
We solve the continuity Eqs.~(\ref{eq1})-(\ref{eq3}) in three dimensions numerically using the values of parameter given in \tref{table2} and virial EOS [see \eref{eq10}]. In \fref{fig05}, we show the data for two sets of values of $C_1$ and $C_2$, the coefficients of viscosity and heat conduction. One set corresponds to the kinetic theory values and the other set corresponds to parameters that we have chosen to fit the data well. In \fref{fig05}, for each of the thermodynamic quantities, the data for four different times collapse onto one curve when they are scaled according to Eqs.~(\ref{eq12})-(\ref{eq15}). Unlike in two dimensions, we find that the data obtained by using the kinetic theory results for $C_1$ and $C_2$ show a slight discrepancy with the MD simulations results (data are taken from Ref.~\cite{joy2021shock3d}). However, by choosing different numerical values for $C_1$ and $C_2$, we find an excellent fit. On the other hand, the TvNS solution, denoted by solid black lines, fail to capture the qualitative features for all the thermodynamic quantities except pressure.

\section{\label{sec4-Summary and discussion }Summary and discussion}

To summarize, we resolved the earlier observed mismatch between the theoretical predictions, based on the Euler equations, for the time evolution of a blast wave and results from hard sphere simulations in two and three dimensions. In this paper, we compare the results of direct numerical simulations of the Navier Stokes equation in two and three dimensions with the MD data, and show they match quantitatively. These results are consistent with those found recently for blast waves in one dimension~\cite{ganapa2021blast,chakraborti2021blast}.  We conclude that the order in which the scaling limit is taken is important in the blast problem. In the TvNS solution, the scaling limit is taken first -- heat conduction and viscosity terms being irrelevant -- before the exact solution is found. However, the correct limit to take is to find the solution with non-zero heat conduction and viscosity terms and then take the scaling limit. 

One possible question that immediately arises is whether the TvNS solution will fail to be the correct description in all dimensions or only upto a critical dimension. One way to answer this question would be to  solve the TvNS equations numerically in $d$ dimensions, and check if gradient of temperature is zero at the shock center. If the gradient is zero, one would expect the heat conduction term to be irrelevant and the TvNS solution to be the correct description. This question is a promising area for future research.

To account for steric effects in hard sphere systems, we used the virial EOS to relate pressure to temperature and density. To show that our results are not sensitive to the EOS used, we compare our results with another well-known EOS, the Henderson EOS, in appendix~\ref{appendix2}.
A generic EOS has the form~\cite{henderson1975simple,santos1995accurate},
\begin{align}
p=\rho k_B T Z(\rho), \label{eq17}
\end{align} 
where $Z(\rho)$ is known as compressibility factor of the EOS.  The classic Henderson relation~\cite{barrat2002molecular} for $Z(\rho)$ for mono-disperse elastic hard sphere gas in two dimensions is given by
\begin{align}
Z(\rho)=\frac{128+\pi^2\rho^2D^4}{8(4-\pi\rho D^2)^2}, \label{eq18}
\end{align}
where $D$ is the diameter of the hard sphere.  \Fref{fig07} shows that there is no significant change in any of the  thermodynamic quantities when a different EOS is used.

The TvNS theory has been extended to study problems where there is a continuous input of energy at the shock center~\cite{dokuchaev2002self}. It would be interesting to compare the theoretical predictions for this problem with MD simulations and check whether it is important to keep heat conduction and viscosity terms. 

\begin{acknowledgements}
The simulations were carried out on the supercomputer Nandadevi at The Institute of Mathematical Sciences.
\end{acknowledgements}

\appendix

\section{\label{appendix} Numerical solution of shock propagation in cartesian coordinates }

In this appendix, we solve for the propagation of the shock in two dimensional cartesian coordinates. Unlike spherical coordinates, where both the initial conditions as well as boundary conditions at $r=0$ have to be provided, in cartesian coordinates only the initial conditions are required for determining the solution at any time. Thus, numerically solving the equations in cartesian coordinates allows us to determine the boundary conditions at $r=0$ for the solution in spherical coordinates. The continuity equations for mass, momentum and energy for a mono-atomic gas are given by~\cite{landaubook,whitham2011linear,huang1963statistical}
\begin{eqnarray}
&&\partial_t\rho + \vec{\nabla}.\rho\vec{u} = 0, \label{A1} \\
&&\partial_t(\rho u_i) + \vec{\nabla}.(\rho  \vec{u}(\vec{u}.i)) + \partial_i p = 2(\vec{\nabla}\mu.\vec{\nabla})u_i + 2\mu((\vec{\nabla}.\vec{\nabla})\vec{u})_i, \label{A2}\\
&&\partial_t\Big[\frac{1}{2}\rho u^2 + \rho T\Big] + \vec{\nabla}.\Big[\Big(\frac{1}{2}\rho u^2 + \rho T + p\Big)\vec{u}\Big] = 2\vec{\nabla}\mu.(\vec{u}.\vec{\nabla})\vec{u} + 2\mu\vec{\nabla}.[(\vec{u}.\vec{\nabla})\vec{u}] + \vec{\nabla}.[\lambda \vec{\nabla}T].~~~~\label{A3}
\end{eqnarray}
\begin{figure}
\centering
\includegraphics[width=\columnwidth]{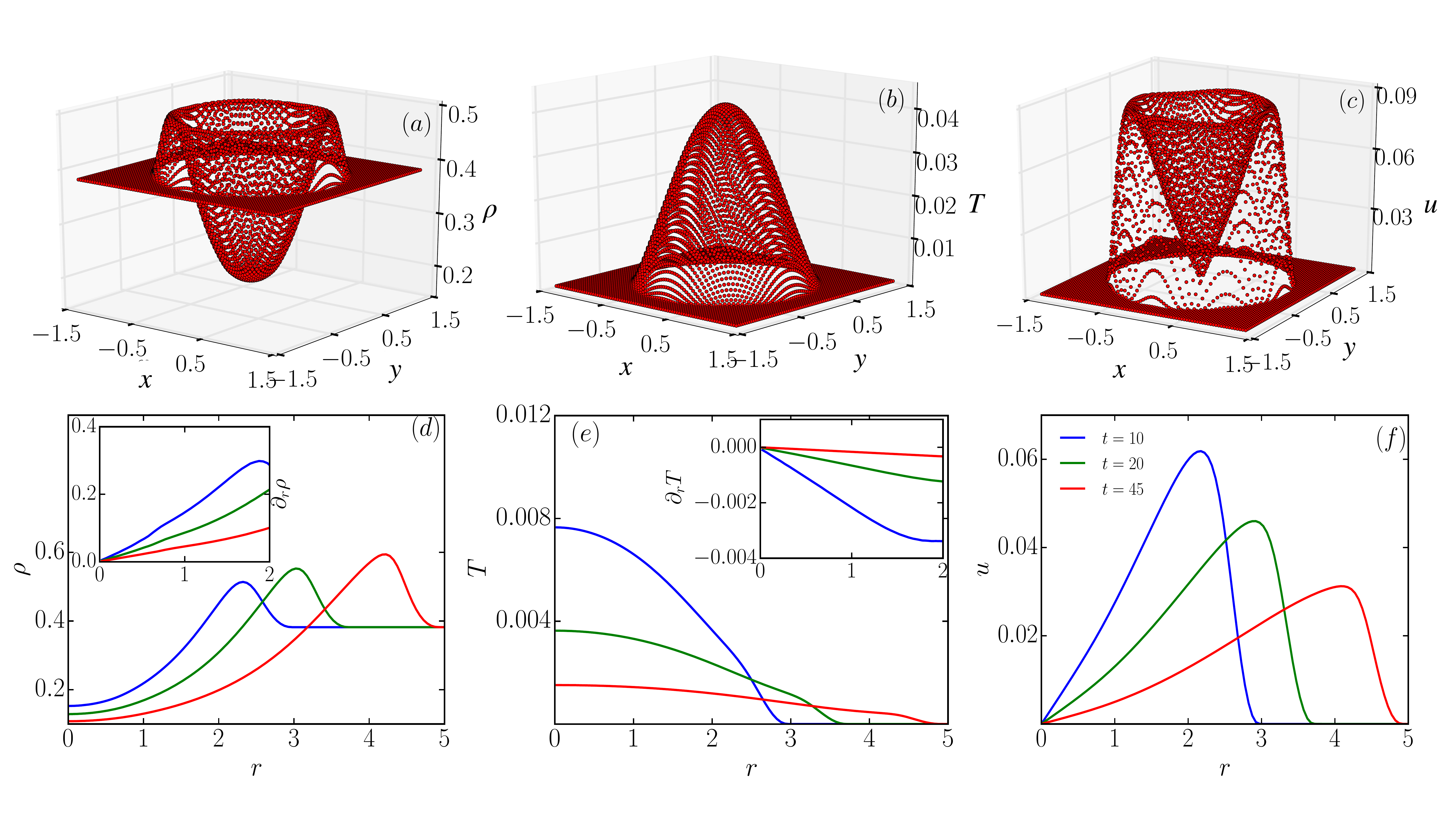}
\caption{The two-dimensional spatial variation of (a) density $\rho$, (b) temperature $T$, and (c) speed $u$, at time $t=2.5$. The corresponding variations with radial distance $r$ are shown in (d)-(f) for three different times. The insets in (d) and (e) show the variation of $\partial_r\rho$ and $\partial_r T$ with $r$. The data justifies the boundary conditions at $r=0$: $\partial_r\rho=0$, $\partial_r T=0$, and $u=0$.} 
\label{fig06}
\end{figure}

We solve Eqs.~(\ref{A1})-(\ref{A3}) in two dimensions using MacCormack method (see \sref{Numerical method}). The pressure is related to temperature and density through the virial EOS given in \eref{eq10}. We choose the same values of the parameters as in spherical coordinate (see \tref{table2}). For spatial discretisation, we use  $dx=dy=dr/\sqrt{2}$. The initial temperature profile is chosen to be a gaussian. The system size is chosen such that the shock does not reach the boundaries during the times that we have considered.

The spatial variation of density, temperature and speed for a single time $t$ is shown in \fref{fig06}(a)-(c) respectively. The corresponding variation with radial distance $r$ are shown in \fref{fig06}(d)-(f). From these data, we can determine the boundary conditions at $r=0$. For all the times shown, $\partial_r \rho=0$ at $r=0$ [see inset of \fref{fig06}(d)]. Also, $\partial_r T=0$ at $r=0$ [see inset of \fref{fig06}(e)]. Finally, $u=0$ at $r=0$  [see \fref{fig06}(f)]. These boundary conditions at $r=0$, $\partial_r\rho=0$, $\partial_r T=0$, and $u=0$, are used for the numerical integration of the continuity equations in radial coordinates (see main text).

\section{\label{appendix2} Comparison between virial EOS and Henderson EOS }

In this appendix we show that our results are not sensitive to the EOS that is used to relate local pressure to local temperature and density. To do so, we compare the results obtained with the virial EOS, presented in the main manuscript, with those obtained from using another commonly used EOS, the Henderson EOS [see Eqs.~(\ref{eq17}) and (\ref{eq18})], and show that the results do not differ much.

In general, the EOS of hard sphere gas is given by \eref{eq17}. There are many proposals for the  compressibility factor $Z(\rho)$~\cite{henderson1975simple,santos1995accurate,barrat2002molecular,mulero2009equation}. Since the general form of $Z(\rho)$ is not known,  we take the commonly used Henderson relation \eref{eq18} for EOS \eref{eq17} to solve the continuity Eqs.~(\ref{eq1})-(\ref{eq3}).

\begin{figure}
\centering
\includegraphics[width=0.8\columnwidth]{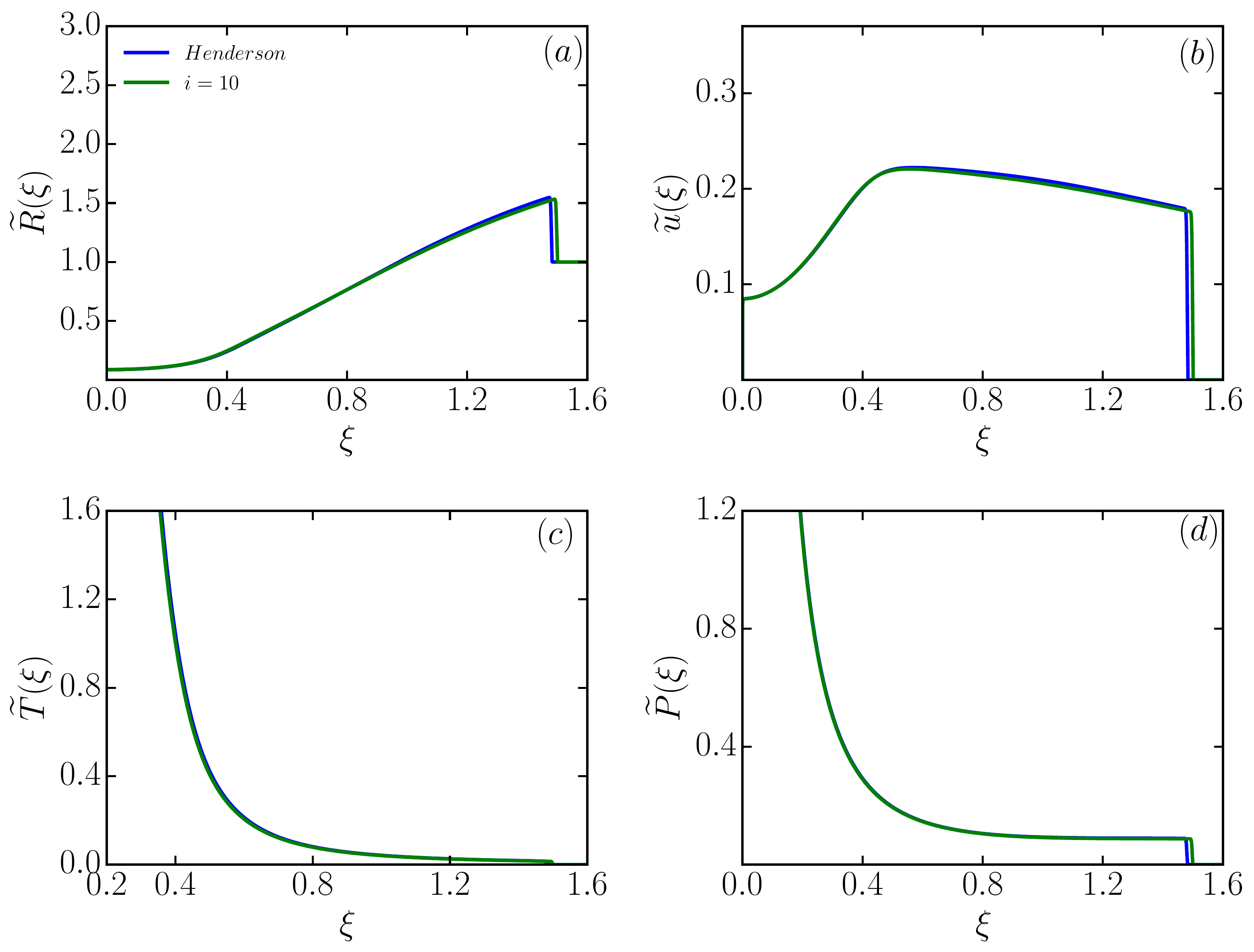}
\caption{The radial variation of non-dimensionalised (a) density, (b) radial velocity, (c) temperature, and (d) pressure  for two different EOS: the virial EOS  truncated at the $10^{th}$ term [see \eref{eq10}] and the Henderson EOS [see Eqs.~(\ref{eq17}) and (\ref{eq18})]. The data are for time  $t=8000$.}
\label{fig07}
\end{figure}

\Fref{fig07} compares the results for density, radial velocity, temperature and pressure for the two EOS. The data for the different EOS cannot be distinguished from each other except for a slight mismatch near the shock front. We conclude that the results presented in the paper are not dependent on the choice of the EOS as long as the EOS describes the equilibrium hard sphere gas well.

\bibliographystyle{spphys}

\end{document}